\documentclass[pdflatex,sn-nature]{sn-jnl}

\usepackage{graphicx}%
\usepackage{multirow}%
\usepackage{amsmath,amssymb,amsfonts}%
\usepackage{amsthm}%
\usepackage{mathrsfs}%
\usepackage[title]{appendix}%
\usepackage{xcolor}%
\usepackage{textcomp}%
\usepackage{manyfoot}%
\usepackage{booktabs}%
\usepackage{algorithm}%
\usepackage{algorithmicx}%
\usepackage{algpseudocode}%
\usepackage{listings}%

\theoremstyle{thmstyleone}%
\theoremstyle{thmstyletwo}%

\theoremstyle{thmstylethree}%

\begin{document}

\title[Dynamic control of ferroic domain patterns by thermal quenching]{Dynamic control of ferroic domain patterns by thermal quenching}


\author*[1]{\fnm{Jan Gerrit} \sur{Horstmann}}\email{jan-gerrit.horstmann@mat.ethz.ch}

\author[1]{\fnm{Ehsan} \sur{Hassanpour}}

\author[1]{\fnm{Yannik} \sur{Zemp}}

\author[1]{\fnm{Thomas} \sur{Lottermoser}}

\author[2]{\fnm{Mads C.} \sur{Weber}}

\author[1]{\fnm{Manfred} \sur{Fiebig}}

\affil*[1]{\orgdiv{Department of Materials}, \orgname{ETH Zurich}, \orgaddress{\street{Vladimir-Prelog-Weg 4}, \city{Zurich}, \postcode{8093}, \country{Switzerland}}}

\affil[2]{\orgdiv{Institut des Molécules et Matériaux du Mans}, \orgname{UMR 6283 CNRS, Le Mans Université}, \orgaddress{\city{Le Mans}, \postcode{72085}, \country{France}}}


\abstract{Controlling the domain structure of ferroic materials is key to manipulating their functionality. Typically, quasi-static electric, magnetic, or strain fields are exploited to transform or pole ferroic domains. In contrast, metallurgy makes use of fast thermal quenches across phase transitions to create new functional states and domain structures. This approach employs the rapid temporal evolution of systems far from equilibrium to overcome the constraints imposed by comparably slow interactions. However, guiding the nonequilibrium evolution of domains towards otherwise inaccessible configurations remains largely unexplored in ferroics. Here, we harness thermal quenches to exert control over a ferroic domain pattern. Cooling at variable speed triggers transitions between two ferroic phases in a rare-earth orthoferrite, with transient domain evolution enabling the selection of the final domain pattern. Specifically, by tuning the quench rate, we can either generate the intrinsic domain structure of the low-temperature phase or transfer the original pattern of the high-temperature phase—creating a hidden metastable domain configuration inaccessible at thermal equilibrium. 
Real-time imaging during rapid quenching reveals two distinct time scales governing domain evolution: a fast fragmentation phase, followed by a slower relaxation towards a new pattern or back to the original one. This dynamic control of domain configurations, alongside external fields, strain engineering, and all-optical switching, offers a novel approach for actively manipulating ferroic order.}

\maketitle

Thermal quenching of materials is ubiquitous in many areas of research and technology. In traditional sword making, for instance, rapidly cooling steel from its high-temperature austenitic state results in the transition to a metastable martensitic phase with superior mechanical properties~\cite{totten_handbook_1993,liscic_theory_1992}. More recently, thermal quenches have been applied to quantum materials, producing phases with exotic electronic characteristics~\cite{de_la_torre_dynamic_2024}. As an overarching theme, these approaches aim to modify the microscopic structure of materials by rapidly changing temperature, yielding final states with novel mechanical~\cite{totten_handbook_1993,liscic_theory_1992}, electrical~\cite{de_la_torre_dynamic_2024,paruch_thermal_2012} or magnetic properties~\cite{eggebrecht_light-induced_2017,li_laser_2020,buttner_observation_2021,griffin_scaling_2012}.\\
\\
Ferroics, with their variety of functional domain structures~\cite{meier_domains_2021,evans_domains_2020,fiebig_evolution_2016}, represent particularly promising targets for quenching schemes. Yet, despite their broad impact on materials science, thermal quenches remain underexplored for shaping ferroic order. For instance, the domain structure of a ferroic is intricately connected to competing long- and short-range interactions within its parent phase~\cite{hubert_magnetic_1998,evans_domains_2020,seul_domain_1995,grunebohm_interplay_2021}. Phase transitions between differently ordered states involve changes in these interactions and are therefore typically accompanied by irreversible modifications of the domain pattern~\cite{hubert_magnetic_1998,portmann_inverse_2003,warmuth_domain_2018}. Changing the phase while maintaining the domain pattern would allow the transfer of mesoscopic functionality to a state with new microscopic interactions or additional coexisting orders. However, it remains to be shown if and how nonequilibrium quenches can be harnessed to control ferroic domain patterns across phase transitions between physically distinct states.\\
\\
Here, we demonstrate the creation of a selected type of domain structure in the multiferroic phase of a rare-earth orthoferrite enabled by thermal quenches. Using laser illumination~\cite{novakovic-marinkovic_stripes_2020,buttner_observation_2021,eggebrecht_light-induced_2017}, the material is heated and rapidly cooled between two magnetically ordered phases with distinct equilibrium domain patterns. We capture the process in real time via Faraday imaging at kilohertz frame rates. Adjusting the quench rate allows switching between phases while either changing the original domain pattern or maintaining it. Following the real-space domain dynamics throughout the quench process reveals how spin reorientation and nonequilibrium domain evolution facilitate control over the final state domain configuration.

\subsection*{Ferroic phases and domains in DTFO}
\begin{figure}[ht!]
    \centering
    \includegraphics[width=0.685\linewidth]{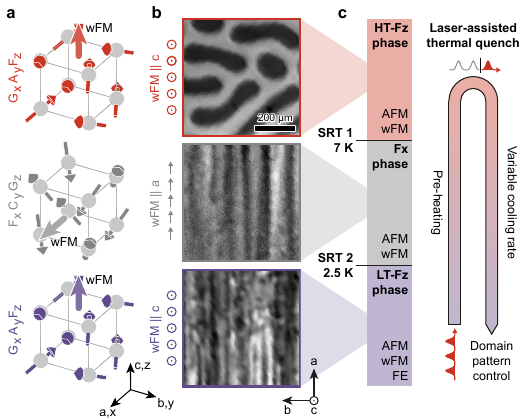}
    \caption{\textbf{$\mid$ Magnetic structure and spin reorientation transitions in DTFO.} \textbf{a}, Temperature-dependent atomic-scale magnetic structure of DTFO in Bertaut notation. Red, gray, and violet arrows, site specific spin orientation; Light red, light gray, and light violet arrows, orientation of the weak ferromagnetic moment. \textbf{b}, Faraday images of the domain pattern at $T_1>T_{\text{SRT1}}$ (top), $T_{\text{SRT1}}>T_2>T_{\text{SRT2}}$ (middle), and $T_3<T_{\text{SRT2}}$ (bottom). To capture the domain image in the $F_x$ phase the sample was rotated by 45 degrees around the $b$-axis to achieve Faraday contrast. \textbf{c}, Temperature-dependent phase diagram of DTFO and laser-assisted thermal quenching scheme. Light red, HT-$F_z$ phase; light gray, $F_x$ phase; light violet, $\text{LT-}F_z$ phase.}
    \label{fig:1}
\end{figure}

\noindent As a model system, we study single crystals of multiferroic $\text{Dy}_{0.7}\text{Tb}_{0.3}\text{FeO}_3$ (DTFO), a rare-earth orthoferrite known for its coexisting, coupled ferroic orders and its numerous magnetic phase transitions~\cite{tokunaga_electric-field-induced_2012,hassanpour_interconversion_2021,hassanpour_magnetoelectric_2022}. In DTFO, the interplay of multiple order parameters and closely spaced phase transitions within a narrow temperature range provides extensive flexibility for tuning and switching between distinct spin and charge configurations~\cite{tokunaga_electric-field-induced_2012,hassanpour_magnetoelectric_2022}. Below $T<T_{\text{Neél}}=653\,\text{K}$, the Fe $3d$ spins of the material arrange in a $G_xA_yF_z$ structure (in Bertaut notation, see Fig.\,\ref{fig:1}a) with a dominant $G_x$-type antiferromagnetic (AFM) order (see Fig.\,\ref{fig:1}a (top))~\cite{tokunaga_electric-field-induced_2012}. Spin-canting of the Fe spins due to the Dzyaloshinskii-Moriya-type interaction induces a weak ferromagnetic (wFM) moment $F_z$ parallel to the $c$-axis of the crystal. Below a critical temperature $T_{\text{SRT1}}=7\,\text{K}$, the system undergoes a spin-reorientation transition (SRT) from a $G_xA_yF_z$ to a $F_xC_yG_z$ structure, with the wFM moment $F_x$ oriented in the sample plane along the $a$-axis (Fig.\,\ref{fig:1}a (middle)). Further lowering the temperature to $T_{\text{SRT2}} < 2.5\,\text{K}$ drives a second SRT back to the $F_z$ structure ($\text{wFM}\parallel c$, Fig.\,\ref{fig:1}a (bottom)). This SRT occurs concomitant with the ordering of the rare-earth moments (Dy, Tb) whose interaction with the Fe sublattice induces the additional ferroelectric (FE) order of the low-temperature (LT) multiferroic $F_z$ phase~\cite{tokunaga_electric-field-induced_2012,hassanpour_magnetoelectric_2022} (see phase diagram in Fig.\,\ref{fig:1}c).\\
\\
To demonstrate dynamic domain-pattern control, we probe the magnetic order in DTFO through its wFM component, which is accessible via linear magneto-optical imaging exploiting the Faraday effect. Figure\,1b shows Faraday images of the same sample area at $T_1>T_{\text{SRT1}}$, $T_{\text{SRT1}}>T_2>T_{\text{SRT2}}$, and $T_3<T_{\text{SRT2}}$. While the high-temperature (HT) $F_z$ phase shows archetypal maze structures with an average domain width of $\sim 80\,\mu\text{m}$, the $F_x$ phase at intermediate temperature exhibits stripe domains oriented along the $a$-axis of the crystal. For cooling under equilibrium conditions, the domain pattern of the LT-$F_z$ phase assumes a similar pattern with stripes along $a$ and additional small-scale modulations attributed to the emergent rare-earth and ferroelectric orders~\cite{tokunaga_electric-field-induced_2012,hassanpour_magnetoelectric_2022}.

\subsection*{Quench-induced domain-pattern transfer}
\begin{figure}[ht!]
    \centering
    \makebox[0pt]{\includegraphics[width=1.37\linewidth]{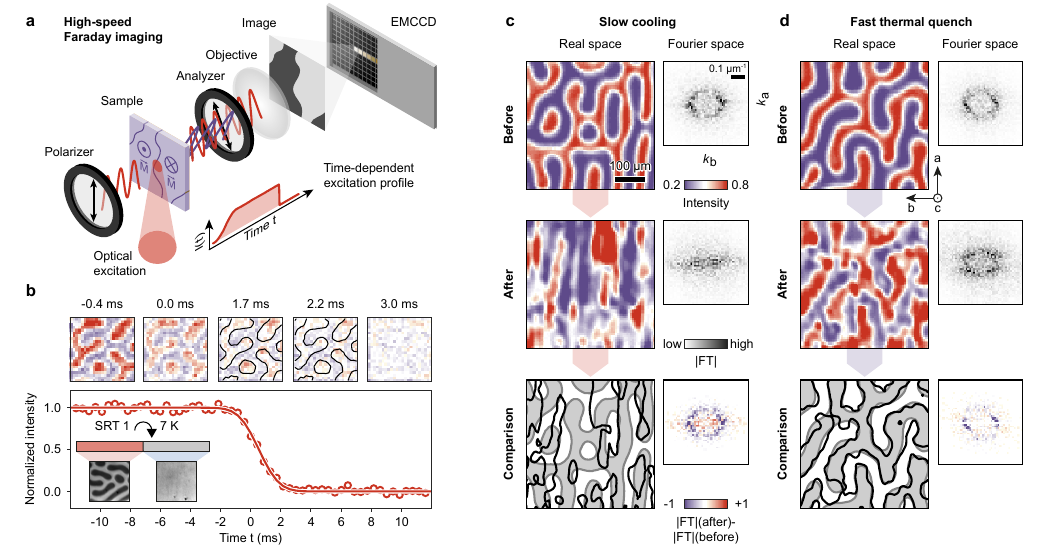}}
    \caption{\textbf{$\mid$ Real-time Faraday imaging of domain dynamics and domain-pattern transfer.} \textbf{a}, Experimental setup for time-resolved Faraday imaging at kHz frame rates. \textbf{b}, Snapshots of the wFM domain evolution across SRT1. Black contours in images `1.7 ms' and `2.2 ms', show original domain pattern before the transition for comparison. \textbf{c}, Domain patterns and corresponding 2D FTs recorded before (top) and after (middle) slow cooling between the $\text{HT-}F_z$ and the $\text{LT-}F_z$ phase (quench rate $\Gamma = 0.4\,\text{K}\,\text{s}^{-1}$). (Bottom) Direct comparison of domain patterns recorded before and after the transition. Light gray contours, before; black contours, after. \textbf{d}, Domain patterns and corresponding 2D FTs recorded before (top) and after (middle) fast thermal quench from the $\text{HT-}F_z$ into the $\text{LT-}F_z$ phase ($\Gamma = 300\,\text{K}\,\text{s}^{-1}$). (Bottom) Direct comparison of domain patterns recorded before (light gray contours) and after (black contours) the transition.}
    \label{fig:2}
\end{figure}

\begin{figure}[ht!]
    \centering
    \includegraphics[width=0.685\linewidth]{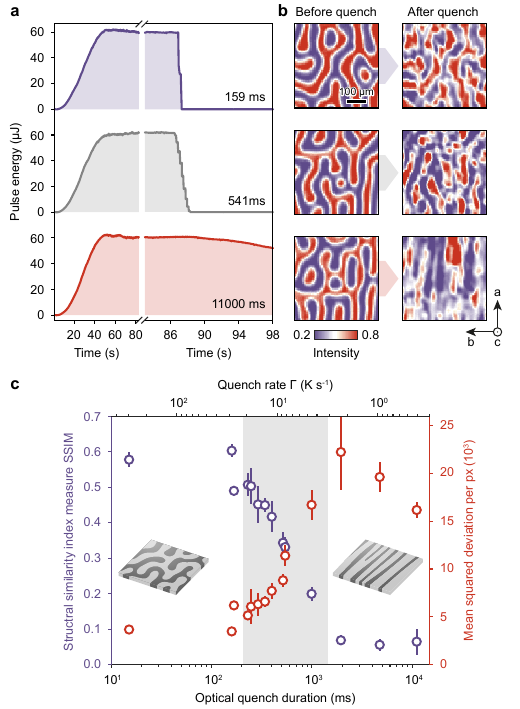}
    \caption{\textbf{$\mid$ Quench-induced magnetic domain-pattern transfer.} \textbf{a}, Time-dependent laser-excitation profiles of three selected thermal quenches with quench durations of $159\,\text{ms}$ (top, violet), $541\,\text{ms}$ (middle, gray), and $11000\,\text{ms}$ (bottom, red). \textbf{b}, Corresponding Faraday images of the same sample area recorded before (left) and after (right) the thermal quenches. \textbf{c}, SSIM (violet circles and axis) and MSD (red circles and axis) of images recorded before and after thermal quenches as a function of the quench duration. Each data point represents the mean value of five quench measurements. Error bars, mean squared error. Light gray area, critical range of optical quench duration ($\Delta t_{\text{q}} = 250-1000\,\text{ms}$).}
    \label{fig:3}
\end{figure}

\noindent The separation of HT- and LT-$F_z$ phases by the intermediate $F_x$ phase within a narrow in temperature range ($T_{\text{SRT1}}-T_{\text{SRT2}}<5$\,K) indicates a delicate energy landscape characterized by pronounced phase competition. This phase competition governs both the microscopic spin orientation and the mesoscopic domain structure in DTFO. Fast changes of the energy landscape may drive the system out of equilibrium, potentially providing a complementary control handle for domain shaping~\cite{de_la_torre_colloquium_2021,buttner_observation_2021,de_la_torre_dynamic_2024}. While rapidly varying magnetic fields are technically challenging, the combination of laser illumination with cryogenic cooling can induce positive and negative temperature changes at very high rates. Here, we investigate the domain dynamics throughout the $\text{HT-}F_z\rightarrow F_x \rightarrow\text{LT-}F_z$ transition after such optically assisted thermal quenches (Fig.\,\ref{fig:1}c, right).\\
\\
We use a train of femtosecond laser pulses to slowly heat the sample from its $\text{LT-}F_z$ state below $2.5\,\text{K}$ into the $\text{HT-}F_z$ phase above $7\,\text{K}$, followed by subsequent cooling at variable rates via time-dependent attenuation of the laser intensity. To monitor the wFM order before, during, and after the thermal quenches, we combine laser heating with high-speed Faraday imaging at a temporal resolution of $\Delta t < 500\,\mu\text{s}$ (Fig.\,\ref{fig:2}a, for detailed information on the technique, see Methods Section). In Faraday imaging, domain selectivity arises from the light's linear polarization plane rotating in proportion to the medium's magnetization parallel to the propagation direction of the light~\cite{mccord_progress_2015}. Selecting the clockwise or anticlockwise rotated field component for imaging creates an intensity contrast between wFM domains with up- or down-magnetization. The sample plane, which in our case coincides with the $c$-plane of the crystal, is imaged onto an electron-multiplying charge-coupled device (EMCCD). This enables image acquisition at frame rates above 1\,kHz with a high signal-to-noise ratio. The laser intensity on the sample is controlled using a half-wave plate on a fast rotation stage and a subsequent polarizer, enabling the programming of time-dependent temperature profiles, synchronized with image acquisition.\\
\\
We showcase the capabilities of high-speed Faraday imaging by resolving the spin-reorientation during the $\text{HT-}F_z\rightarrow F_x$ transition (Fig.\,\ref{fig:2}b) driven far from equilibrium at a quench rate of $300\,\text{K}\,\text{s}^{-1}$. We observe a loss of Faraday contrast between domains of opposite magnetization within $1\,\text{ms}$, which we attribute to the rotation of the wFM moments from their original out-of-plane to an in-plane orientation upon the $\text{HT-}F_z\rightarrow F_x$ transition~\cite{tokunaga_electric-field-induced_2012}. The time scale of this reorientation is determined by the cooling speed across the phase transition. Snapshots of the domain pattern at different time delays across the SRT (Fig.\,\ref{fig:2}b, top) show that the maze-domain pattern remains stable, while its contrast vanishes (compare residual intensities in snapshots `1.7 ms' and `2.2 ms' with initial pattern (black contours)). This suggests that during fast cooling, the orientation of the local wFM moments and the global domain configuration evolve on different time scales. Such behavior provides initial evidence that the nonequilibrium transfer of domain patterns across the SRT and into a target phase may be feasible.\\
\\
To explore the potential of such quenching schemes for domain-pattern control, we investigate the wFM domain structures across both phase transitions, that is, from the $\text{HT-}F_z$ phase, passing the intermediate $F_x$ phase, and into the $\text{LT-}F_z$ phase. We image the domain patterns before and after: first, for a slow cooling process allowing the domain pattern to adjust to the spin rotations (Fig.\,\ref{fig:2}c). Second, for a fast thermal quench driving the system far from equilibrium (Fig.\,\ref{fig:2}d). Starting from maze patterns in the $\text{HT-}F_z$ phase (Fig.\,\ref{fig:2}c, top), slow cooling (cooling rate $\Gamma = 0.4\,\text{K}\,\text{s}^{-1}$) results in stripe patterns in the $\text{LT-}F_z$ phase (Fig.\,\ref{fig:2}c, middle) with no apparent correlation between initial and final domain patterns. The emerging anisotropy of the wFM domains is equally reflected in the vanishing of the circular feature in two-dimensional Fourier transforms (FTs) of the $\text{LT-}F_z$ pattern (compare FTs in Fig.\,\ref{fig:2}c, top/middle/bottom). For fast cooling (quench rate $\Gamma = 300\,\text{K}\,\text{s}^{-1}$), however, we find a markedly different behavior. Most importantly, we observe a nearly complete transfer of the initial maze pattern of the $\text{HT-}F_z$ phase to the $\text{LT-}F_z$ phase (see direct comparison of domain contours and Fourier transforms of both phases in Fig.\,\ref{fig:2}d, bottom).\\
\\
The above results demonstrate the ability to select between two distinct types of domain patterns, namely maze and stripe domains, in the targeted $\text{LT-}F_z$ phase. Notably, the LT maze pattern is inaccessible through an equilibrium pathway. The cooling speed represents the control parameter, which implies a critical quench rate $\Gamma_{\text{crit}}$ governing the maze-domain-pattern transfer. To determine $\Gamma_{\text{crit}}$, we perform a series of quench experiments at predefined quench durations $\Delta t_{\text{q}}$ (Fig.\,\ref{fig:3}a) and analyze the similarity between initial- and final-state domain patterns (Fig.\,\ref{fig:3}b). A comparison of images recorded before (left) and after (right) quenching highlights the gradual decrease (increase) of visual correlation for slower (faster) cooling.\\
\\
We evaluate the similarity of initial and final domain structures in two metrics. The mean squared deviation (MSD) compares intensity deviations between the images in a pixel-by-pixel manner, whereas the structural similarity index (SSIM)~\cite{wang_image_2004} additionally accounts for correlations on larger length scales (for details, see Methods Section). Analyzing the MSD and SSIM as a function of the optical quench duration (Fig.\,\ref{fig:3}c) reveals a critical range between $\Delta t_{\text{q}}=250-1000\,\text{ms}$ around a threshold value of $\Delta t_{\text{q}}^{\text{mean}}=625\,\text{ms}$ governing the domain-pattern transfer. The latter value corresponds to a quench rate $\Gamma_{\text{crit}} = 7.2\,\text{K}\,\text{s}^{-1}$. For $\Gamma\ll\Gamma_{\text{crit}}$, the $\text{LT-}F_z$ phase exhibits stripe domains. For $\Gamma\gg\Gamma_{\text{crit}}$, on the other hand, the original maze-domain pattern is transferred to the $\text{LT-}F_z$ phase.

\subsection*{Nonequilibrium domain evolution}

\begin{figure}[ht!]
    \centering
    \includegraphics[width=0.685\linewidth]{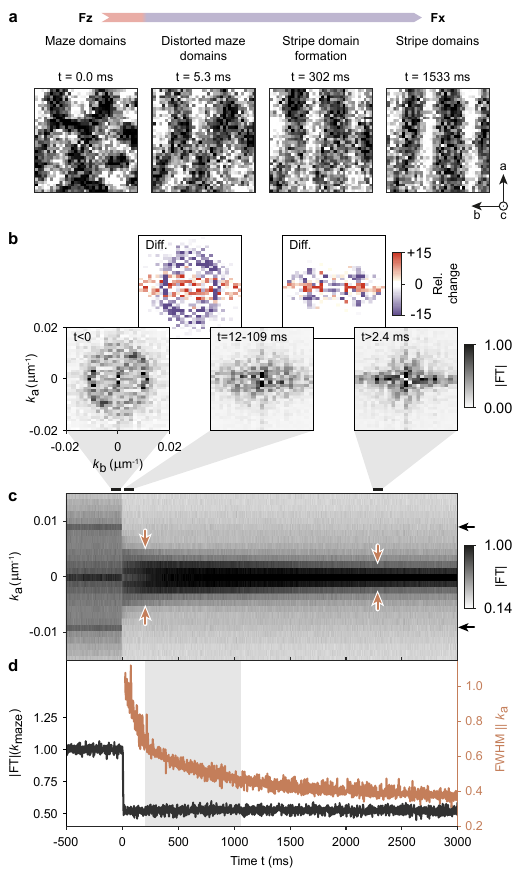}
    \caption{\textbf{$\mid$ Time-scales of spin-reorientation and domain dynamics.} \textbf{a}, Snapshots of initial domain dynamics and stripe pattern formation in the $F_x$ phase. \textbf{b}, 2D FTs of domain patterns at three selected times during the quench between the $\text{HT-}F_z$ and the $F_x$ phase (bottom) and difference images (top). Bottom row: Left, before the $\text{HT-}F_z\rightarrow F_x$ SRT ($t<0\,$ms, $|\text{FT}|_{Fz}$). Middle, directly after the SRT ($12\,\text{ms}<t<109\,\text{ms}$, $|\text{FT}|^{\text{initial}}_{Fx}$). Right, at later times ($t>2.4\,\text{s}$, $|\text{FT}|^{\text{final}}_{Fx}$). Top row: Left, difference image of 2D FTs recorded before and directly after the SRT, $(|\text{FT}|^{\text{initial}}_{Fx}-|\text{FT}|_{Fz})/\langle |\text{FT}|_{Fz}\rangle$. Right, difference image of 2D FTs recorded directly after the SRT and at later times, $(|\text{FT}|^{\text{final}}_{Fx}-|\text{FT}|^{\text{initial}}_{Fx})/\langle |\text{FT}|^{\text{initial}}_{Fx}\rangle$. \textbf{c}, FTs integrated along $k_b$ as a function of time $t$ relative to SRT1. Black and brown arrows mark the features which are further analyzed in \textbf{d}. For details on the Fourier component at $k_a=0$ see Fig.\,S\ref{fig:S4}. \textbf{d}, Intensity of ring feature ($|\text{FT}|(k_{\text{maze}})$, black) in \textbf{b} and full width half maximum of the Fourier amplitude distribution along $k_a$ ($\text{FWHM}\parallel k_{\text{a}}$, brown) as a function of time.}
    \label{fig:4}
\end{figure}

We next investigate the mechanism underlying the observed dynamic domain control, focusing on how nonequilibrium domain evolution across both SRTs and the intermediate $F_x$ phase enables selecting the final domain configuration. To directly access domain dynamics in the interim $F_x$ phase, we perform time-resolved Faraday imaging on tilted samples. By doing so, we achieve Faraday contrast in both $F_z$ and $F_x$ phases despite the in-plane net-magnetization of the $F_x$ phase.\\
\\
We begin by examining the dynamics across SRT1 and stripe-domain formation in the $F_x$ phase during rapid quenches from $T > 7\,\text{K}$ to $T = 4\,\text{K}$ (Fig.\,\ref{fig:4}a). For quantitative data analysis, we extract the temporal evolution of specific features in the domain pattern from FTs of the time-dependent domain configurations in a Faraday movie (Fig.\,\ref{fig:4}b-d). We find that the quench behavior splits into two processes occurring on vastly different time scales. First, SRT1 causes a rapid distortion of the long-range ordered, isotropic maze domains within tens of milliseconds (compare images recorded at `$t= 0.0\,\text{ms}$' and `$t= 5.3\,\text{ms}$' in Fig.\,\ref{fig:4}a). In the FT, this is reflected in the fast suppression of the circular feature (compare FTs of domain patterns recorded before (`$t< 0.0\,\text{ms}$') and directly after (`$t = 12-109\,\text{ms}$') SRT1 in Fig.\,\ref{fig:4}b, as well as the corresponding difference image). The precise evolution of the intensity of this ring feature evidences that the abrupt domain-pattern change takes place a timescale comparable to SRT1 itself ($|\text{FT}|(k_{\text{maze}})$, black trace in Figure\,\ref{fig:4}d). The distortion of the maze pattern occurs in form of a domain stretching (compression) along the $a$ ($b$) axis of the crystal as the evolution in Fig.\,\ref{fig:4}a shows. This is also apparent from the emerging elongated feature along $k_b$ (see `$t = 12-109\,\text{ms}$' and left difference image in Fig.\,\ref{fig:4}b). In other words, the well-defined maze structure breaks up into smaller stripe-like domains forming a pointillistic image of the original maze pattern. Second, the distorted maze pattern evolves into a regular stripe pattern on a longer time scale of hundreds of milliseconds (see `$t= 302\,\text{ms}$' and `$t= 1553\,\text{ms}$' in Fig.\,\ref{fig:4}a). In the FT, this manifests in a narrowing of the elongated feature in $k_a$-direction over time (`$t= 2.4\,\text{s}$' in Fig.\,\ref{fig:4}b, corresponding difference image, and brown trace in Fig.\,\ref{fig:4}d). Importantly, the characteristic time scale of stripe-domain formation is in quantitative agreement with the critical quench time $\Delta t_{\text{q}}^{\text{mean}}=625\,\text{ms}$ for the domain pattern transfer into the $\text{LT-}F_z$ phase (compare light gray areas in Fig.\,\ref{fig:3}c and Fig.\,\ref{fig:4}d). This points to a critical role of domain evolution in the interim $F_x$ phase in controlling the domain pattern of the targeted $\text{LT-}F_z$ phase.\\
\\
To complete our picture of domain-pattern control, we now focus on the domain evolution following the SRT2 transition into the $\text{LT-}F_z$ phase below $2.5\,\text{K}$. To this end, we record fast real-time Faraday movies of the domain dynamics throughout the entire nonequilibrium quench process from the $\text{HT-}F_z$ through the $F_x$ to the $\text{LT-}F_z$ phase (Fig.\,\ref{fig:5}a). As expected, the original maze domain pattern of the $\text{HT-}F_z$ phase (`$t<0\,\text{ms}$') is distorted in the transition to the $F_x$ phase (`$t=0-17\,\text{ms}$') and forms a new transient configuration (`$t=30-52\,\text{ms}$', compare features highlighted by red circles and rectangles in first three images of Fig.\,\ref{fig:5}a). More specifically, individual domains rotate towards the $a$-axis of the crystal (see co-rotating red rectangles in (1), (2), and (3)), in line with the preferential orientation of stripe domains in the $F_x$ phase. The contraction of adjacent domains further leads to a local enhancement of the Faraday signal (highlighted by stationary red circles in (1), (2), and (3)). Remarkably, with the transition into the $\text{LT-}F_z$ phase, the distorted domains are not simply frozen in, but actively evolve back towards the original structure on a time scale of multiple seconds ((4) `$t=3400\,\text{ms}$', (5) `$t=7700\,\text{ms}$'). By contrast, in the adiabatic transition, this back-rotation towards the original maze structure is absent, and stripe domains form and ripen instead.

\subsection*{Discussion}
\begin{figure}[ht!]
    \centering
    \includegraphics[width=0.685\linewidth]{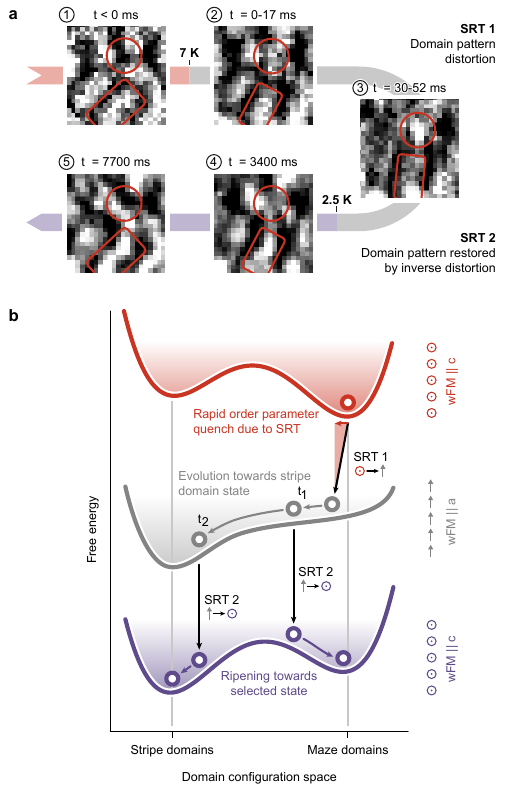}
    \caption{\textbf{$\mid$ Back-relaxation of domains and dynamic free-energy model of domain-pattern control}. \textbf{a}, Domain evolution between $\text{HT-}F_z$ and $\text{LT-}F_z$ phases during a fast quench ($\Gamma = 300\,\text{K}\,\text{s}^{-1}$). Light red, $\text{HT-}F_z$ phase; light gray, $F_x$ phase; light violet, $\text{LT-}F_z$ phase. Red rectangles and circles highlight selected features of the domain structure discussed in the main text. \textbf{b}, Dynamic free-energy model of domain-pattern control. Schematic free-energy surfaces of DTFO in domain configuration space. Red, $\text{HT-}F_z$ phase; gray, $F_x$ phase; violet, $\text{LT-}F_z$ phase. Red, gray, and violet circles represent the momentary position of the system in configuration space.}
    \label{fig:5}
\end{figure}

\begin{figure}[ht!]
    \centering
    \makebox[0pt]{\includegraphics[width=1.37\linewidth]{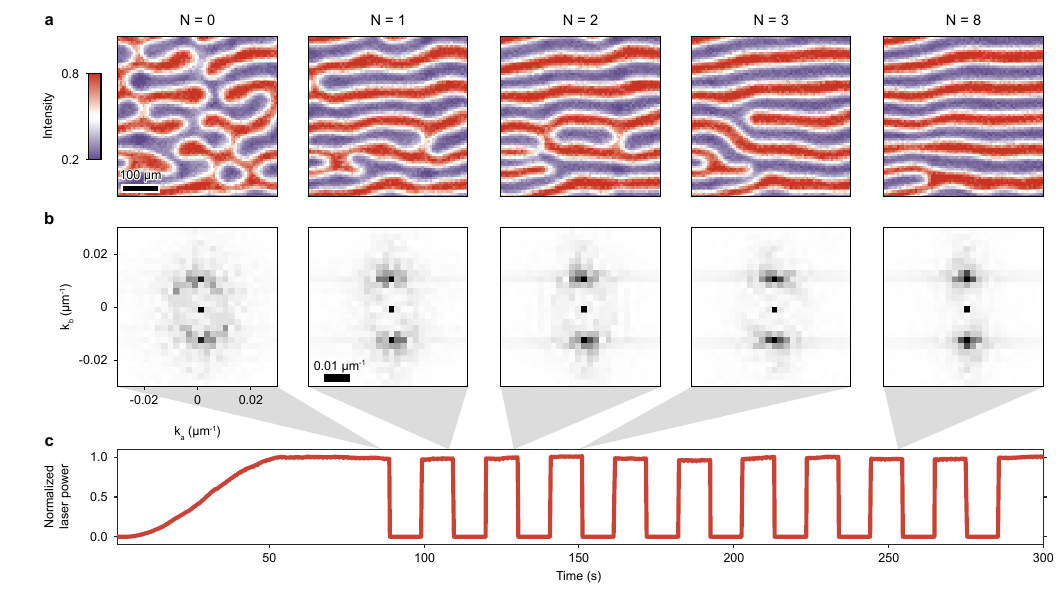}}
    \caption{\textbf{$\mid$ S2: Back and forth quenches between LT- and HT-$F_z$ phases.} \textbf{a}, Faraday images of the sample in the $\text{HT-}F_z$ phase recorded after $N$ fast laser-assisted quenches into the $\text{LT-}F_z$ phase and back. With increasing $N$, the initial maze-domain pattern is transformed step-by-step into a stripe-domain pattern due to the repeated passage of the intermediate $F_x$ phase. Because of the preservation of the domain pattern between the two phases during quenches, the creation or annihilation of individual topological defects can be observed. The quench-induced stripe pattern in the $\text{HT-}F_z$ phase is stable for many hours. Maze domains can only be regained by heating the sample up to $\sim100\,\text{K}$ and slowly cooling down again. \textbf{b}, Two-dimensional Fourier transforms of the respective images in \textbf{a} highlight the stepwise loss of isotropy of the domain pattern due to stripe formation. \textbf{c}, Time-dependent laser power on the sample during the successive heating/cooling cycles. Based on these results and the data presented in Fig.\,\ref{fig:2}, we conclude that both the $\text{HT-}F_z$ and the $\text{LT-}F_z$ phases exhibit metastable maze domain and stripe domain configurations. Whereas maze domains are energetically favorable in the $\text{HT-}F_z$, stripe domains likely represent the lower-energy configuration in the $\text{LT-}F_z$ phase. Consequently, stripe (maze) domains represent a metastable configuration in the $\text{HT-}F_z$ ($\text{LT-}F_z$) phase.}
    \label{fig:S2}
\end{figure}

The domain-pattern transfer at high quench rates, the role of domain evolution in the intermediate $F_x$ state, and the back-rotation of quenched domains towards their initial configuration in the targeted $\text{LT-}F_z$ phase can be understood within a dynamic free-energy model (Fig.\,\ref{fig:5}b)~\cite{landau_theory_1937,toledano_landau_1987}. For thin ferromagnetic films with perpendicular orientation of the magnetization with respect to the surface, it is known that bubble, maze, and stripe domains patterns can coexist at zero field as metastable states in configuration space, separated by free-energy barriers~\cite{hubert_magnetic_1998}. External fields, seeding by structural defects, or secondary anisotropy contributions can tilt the balance towards a particular type of domain configuration. With respect to our system, fast cooling (and equally fast heating, see Fig.\,S\ref{fig:S2}) between the two $F_z$ phases evidences the bi-stability of different domain patterns. More specifically, wFM stripe and maze domains represent (meta-)stable states in both phases, suggesting a description of initial ($\text{HT-}F_z$) and final ($\text{LT-}F_z$) states in terms of double-well Landau-free-energy surfaces.\\
\\
In this picture, starting from the maze-domain configuration in the $\text{HT-}F_z$ phase, the first SRT induces a rapid initial displacement in configuration space (horizontal red arrow in Fig.\,\ref{fig:5}b) and transfers the system into the $F_x$ state. For this $F_x$ phase, the observed formation of stripe domains point to a free-energy surface exhibiting only one stable state, namely the stripe-domain configuration. Consequently, the system evolves towards the new stable minimum until the second SRT restores the free-energy surface to its double-well form. In this scenario, the time spent in the $F_x$ phase becomes a decisive parameter: For short quench times ($t_1$), the system starts to evolve towards the stripe configuration in $F_x$, but is driven back into the maze configuration on the free-energy surface of the $\text{HT-}F_z$ phase. Long quench times ($t_2$), on the other hand, facilitate the formation of stripes in the intermediate $F_x$ phase, which are then transferred to the multiferroic $\text{LT-}F_z$ phase. In both cases, gradients in the free-energy landscape force an evolution of transferred domain structures towards the local minimum in configuration space in the $\text{LT-}F_z$ phase, as observed by time-resolved Faraday imaging (compare Fig.\,\ref{fig:5}a).\\
\\
We note that while the quench-induced maze-domain structure represents a long-lived metastable state in the $\text{LT-}F_z$ phase (lifetime $\tau_{\text{maze}}^{\text{LT-}F_z}\gg 1\,\text{h}$), it can only be accessed via a nonequilibrium pathway. This suggests a rather general anisotropy of the system that favors the stripe-domain configuration as the ground state of the wFM order. The quench-induced maze-domain state, by contrast, has to be forcefully imprinted. Future studies could explore the impact of this hidden domain state on the domain formation in the coexisting rare-earth and ferroelectric phases. 
\\ \\
Beyond the model system of DTFO, we propose that adapting thermal quenching schemes to other classes of ferroics or correlated materials with electronic phase separation could unlock access to previously unknown or hidden (domain) configurations of matter, potentially without the immediate need for ultrafast optical excitation~\cite{de_la_torre_dynamic_2024}. In this, fast real-time imaging~\cite{klose_coherent_2023,kronseder_dipolar-energy-activated_2013,kronseder_real-time_2015} of irreversible switching events and fluctuations, as demonstrated here, serves as an essential tool for exploring, understanding, and controlling transitions into these novel phases.

\subsection*{Conclusion}

In summary, we have demonstrated dynamic control over the domain pattern of a multiferroic phase by functionalizing phase transitions as effective switches between distinct types of domains. Our results show that fast, nonequilibrium quenches and the resulting domain dynamics not only generate new domain structures, but also enable the transfer of existing domain patterns into a target phase. 
Combined with external fields~\cite{leo_magnetoelectric_2018,haykal_antiferromagnetic_2020,chesnel_shaping_2016,schlom_elastic_2014}, all-optical control schemes~\cite{buttner_observation_2021,zalewski_ultrafast_2024,berruto_laser-induced_2018,eggebrecht_light-induced_2017,kirilyuk_ultrafast_2010,stanciu_all-optical_2007,manz_reversible_2016}, or the mutual coupling of different orders~\cite{zhao_electrical_2006,hassanpour_magnetoelectric_2022,leo_magnetoelectric_2018}, our approach offers a powerful means for switching functional domain configurations in technologically relevant states. This refined manipulation of domains could significantly impact the development of devices that leverage microscopic ferroic heterogeneity~\cite{parkin_magnetic_2008}.

\newpage

\backmatter

\bibliography{DTFO_v4}

\section*{Methods}\label{secA1}

\textbf{Sample preparation.} DTFO single crystals were grown by Y. Tokunaga and Y. Tokura at the RIKEN Center for Emergent Matter Science (CEMS). The crystals were cut into thin sheets of $\sim10\,\text{mm}^2$ expanding in the $c$-plane and lapped down to a thickness of $60\,\mu\text{m}$. This results in an optical transmission of $\sim 10\,\%$ in the visible range. Both sample faces were polished with a silica slurry to minimize diffuse light scattering in the imaging experiments. After mechanical treatment, the crystals were annealed at temperatures $T>700\,^{\circ}\text{C}$ to release mechanical stress introduced by the lapping/polishing process. The samples were mounted on customized holders, the surface normal aligned either at $0\,^{\circ}$ or at $45\,^{\circ}$ with respect to the optical axis (see Supplementary Fig.\,S\ref{fig:S3}). For the imaging experiments, the sample holders were placed in an optical cryostat (Oxford Spectromag) and cooled to base temperatures $T_{\text{base}}$ between 2 and 10$\,\text{K}$.\\
\\
\textbf{Experimental setup.} In time-resolved imaging experiments, the samples were homogeneously illuminated by the collimated output of a LED (Thorlabs M660L4, central wavelength $\lambda_{\text{c}} = 660\,\text{nm}$). The typical LED driving current in experiments ranged from $10-80\,\text{mA}$ corresponding to optical powers of $7.6-56.8\,\mu\text{W}$ distributed over the whole sample area. These values are vanishingly small compared with the pump-beam power. No probe-beam-induced heating was observed during measurements. A camera objective (focal length $f = 200\,\text{mm}$) was used to image the sample onto the camera sensor. Between objective and sensor, the beam passes a Glan-Taylor polarizer on an adjustable rotation mount to generate the Faraday imaging contrast, and is passed through a band-pass filter (central wavelength $\lambda_{\text{c}}=650\,\text{nm}$, bandwidth $\Delta\lambda=40\,\text{nm}$). A schematic of the entire optical setup is shown in Supplementary Fig.\,\ref{fig:1}. As a detector, we employ an electron-multiplying charge coupled device (EMCCD, Andor iXon Ultra 897). In EMCCDs (Electron Multiplying Charge-Coupled Devices), electron signals from the CCD chip are amplified above the read-noise floor by a multiplication register before reaching the readout amplifier. This process enables image capture with high sensitivity at high frame rates. To achieve frame rates of up to 2350 frames per second, we use the camera's `frame transfer' and `crop' modes. An adjustable slit with four independent lamellas masks the CCD chip, leaving a pre-selected sub-area of variable size (typically $256\times256$, $128\times128$, or $64\times64$ pixel, depending on the desired spatial and temporal resolution). This setup allows to store recently recorded images in the masked sections of the chip while the current image forms in the unmasked area.\\
\\
To initiate and control the wFM domain dynamics in the sample, we use the output of a femtosecond laser amplifier (Coherent Legend Elite Duo, central wavelength $\lambda_{\text{c}}=800\,\text{nm}$, maximum pulse energy $E_{\text{p}}^{\text{max}}=60\,\mu\text{J}$, repetition rate $f_{\text{rep}} = 1\,\text{kHz}$, maximum average power $P_{\text{cw}}^{\text{max}}=60\,\text{mW}$) to heat the sample. For this purpose, the pump beam is attenuated by using a half-wave plate mounted in a fast motorized rotation stage (Thorlabs ELL14) and a subsequent, fixed Glan-Taylor polarizer. During measurements, the temporal profile of the optical excitation was monitored and logged by guiding a small fraction of the pump beam onto a photodiode placed after the attenuator. The pump beam was slightly focused onto the back side of the sample by a 400-mm lens, resulting in a spot size of $\sim 500\times500\,\mu\text{m}^2$, which is large compared to the imaged sample area. The maximum excitation fluence was $7.6\,\text{mJ\,cm}^{-2}$ for $P_{\text{cw}}=60\,\text{mW}$. Taking $T_{\text{SRT1}}$ ($T_{\text{SRT2}}$) and $P_{\text{cw}}^{\text{SRT1}}$ ($P_{\text{cw}}^{\text{SRT2}}$) as reference points, we determine the laser-induced temperature increase to $\Delta T_{\text{sample}}/\Delta P_{\text{cw}}=0.17\,\text{K}\,\text{mW}^{-1}$. On this basis, we estimate a maximum temperature increase of $\Delta T=10.3\,\text{K}$ for $P_{\text{cw}}^{\text{max}}=60\,\text{mW}$, corresponding to a maximum sample temperature of $12.7\,\text{K}$. Synchronization between optically-assited thermal quenches and image acquisition was realized by a customized LabView program.\\
\\
\textbf{Data acquisition.} For the time-resolved Faraday-imaging experiment on the SRT at $T_{\text{SRT1}}$ (Fig.\,\ref{fig:2}b), the DTFO sample was heated from its initial $\text{LT-}F_x$ state at a base temperature of $T=4\,\text{K}$ into the $\text{HT-}F_z$ phase by setting increasing averaged laser power to $60\,\text{mW}$. The surface normal of the sample was oriented parallel to the propagation direction of the probe light, maximizing (minimizing) the domain contrast in the $\text{HT-}F_z$ ($F_x$) phase. The laser beam was rapidly attenuated by a fast mechanical shutter (shutter closing time $t_{\text{close}} < 3\,\text{ms}$) and synchronised to the EMCCD camera acquiring images ($64\times64\,\text{pixels}$, $2\times2$ binning, resulting image size after binning: $32\times32\,\text{pixels}$ ) at a frame rate of 2325 Hz (corresponding temporal resolution $\Delta t_{\text{res}} = 430\,\mu\text{s}$). For the measurements on the quench-induced domain-pattern transfer in Fig.\,\ref{fig:2}c,d and Fig.\,\ref{fig:3}a,b, the sample was heated from its initial $\text{LT-}F_z$ state at a base temperature of $T=2.45\,K$ into the $\text{HT-}F_z$ phase by gradually increasing the laser heating power from $0\,\text{mW}$ to $60\,\text{mW}$ over a period of $30\,\text{s}$ (see time-dependent heating trace Fig.\,\ref{fig:3}a) with the attenuator described above. Following the heating step, the sample was held at maximum temperature for $10\,\text{s}$ to stabilize the wFM domain pattern, followed by a thermal quench of predefined duration. Reference images were taken $5\,\text{s}$ before and $5\,\text{s}$ after the quench with typical integration times of $50\,\text{ms}$. To ensure the comparability between measurement runs, all steps of the experiment, that is, heating, annealing, quenching, image acquisition, and the repeated heating were fully automatized. The images shown in Figs.\,\ref{fig:4}a,e were recorded at a frame rate of $569\,\text{Hz}$ ($128\times128\,\text{pixel}$) with $1\times1$ and $2\times2$ binning (in post-processing), respectively. The corresponding temporal resolution was $1.7\,\text{ms}$.\\
\\
\textbf{Data analysis.} We determine the optical quench duration $\Delta t_{q}$ by fitting an error function model to the reference data for the optical excitation collected by a photodiode (see Fig.\,S1 for details of the experimental setup) before, during, and after the laser-assisted quench. To quantitatively analyze the similarity between domain patterns before and after the thermal quenches as a function of $\Delta t_{q}$, we evoke two different metrics. The mean standard deviation (MSD) or mean standard error (MSE), respectively, represents a pixel-by-pixel comparison of image intensities. For images $I^{\text{b}}$ and $I^{\text{a}}$ with $N\times M$ pixels, recorded before and after the quench,
\begin{align}
    \text{MSE}(I^{\text{b}},I^{\text{a}})=\frac{1}{NM}\sum_{i=1}^{N}\sum_{j=1}^{M}\left(I_{ij}^{\text{b}}-I_{ij}^{\text{a}}\right)^2,
\end{align}
where $I_{ij}^{\text{b}}$ ($I_{ij}^{\text{a}}$) is the intensity of the $ij$-th pixel before (after) the quench. The structural similarity index measure (SSIM), on the other hand, compares images in terms of three different parameters: luminosity $l$, contrast $c$, and structure $s$, according to 
\begin{align}
    \text{SSIM}(I^{b},I^{a}) = \left[l(I^{b},I^{a})\right]^{\alpha}\cdot\left[c(I^{b},I^{a})\right]^{\beta}\cdot\left[s(I^{b},I^{a})\right]^{\gamma}.\label{eq:SSIM_long}
\end{align}
The exponents $\alpha$, $\beta$, and $\gamma$ can be chosen freely between 0 and 1 to give more weight to specific contributions. The measures of luminosity, contrast, and structure are defined by
\begin{align}
    l(I^{b},I^{a}) &= \frac{2\mu_b\mu_a+c_1}{\mu_b^2+\mu_a^2+c_1},\;\\
    c(I^{b},I^{a}) &= \frac{2\sigma_b\sigma_a+c_2}{\sigma_b^2+\sigma_a^2+c_2},\\
    s(I^{b},I^{a}) &= \frac{\sigma_ba+c_3}{\sigma_b\sigma_a+c_3}.
\end{align}
Here, $\mu_b$ ($\mu_a$) is the mean intensity of the image before (after) the quench, $\sigma_b^2 = 1/(NM-1)\sum_{i=1}^{N}\sum_{j=1}^{M}(I^b_{ij}-\mu_b)^2$ is the variance of $I^b$ (analogously, $\sigma_a^2$ is the variance of $I^a$), and $\sigma_{ba} = 1/(NM-1)\sum_{i=1}^{N}\sum_{j=1}^{M}(I^b_{ij}-\mu_b)(I^a_{ij}-\mu_a)$ the covariance of $I^b$ and $I^a$. The constants $c_1=(k_1L)^2$, $c_2=(k_2L)^2$, and $c_3=(k_3L)^2$ are used to stabilize the SSIM in case of small denominators. Typically, values of $0.01$ and $0.03$ are chosen for $k_1$ and $k_2$, respectively, whereas $L$ depends on the number of bits per pixel. In our case, $L=2^{16}-1$. For $\alpha = \beta = \gamma = 1$, Eq.\,\ref{eq:SSIM_long} reduces to
\begin{align}
    \text{SSIM}(I^b,I^a) = \frac{(2\mu_b\mu_a+c_1)(2\sigma_{ba}+c_2)}{(\mu_b^2+\mu_a^2+c_1)(\sigma_b^2+\sigma_a^2+c_2)}.\label{eq:SSIM}
\end{align}
Evaluating Eq.\,\ref{eq:SSIM} as a function of the optical quench duration yields the data presented in Fig.\,\ref{fig:3}b.

\section*{Declarations}

\begin{itemize}
\item \textbf{Acknowledgements.} This work was funded by an ETH Postdoctoral Fellowship and a SNSF Postdoctoral Fellowship (TMPFP2 217303). The authors would like to thank M. Trassin, Y. Tokunaga, R. V. Pisarev, and A. Vaterlaus for helpful discussions, as well as J. Hecht and S. Reitz for technical support.
\item \textbf{Author contributions}. The project was conceived by J.G.H. and M.F. Experiments and data analysis were conducted by J.G.H., with contributions from E.H., Y.Z, T.L. and M.C.W. The manuscript was written by J.G.H. All authors discussed the results and commented on the manuscript.
\item \textbf{Competing interests.} The authors declare no competing interests.
\item \textbf{Materials \& Correspondence}. Correspondence and requests for materials should be addressed to J.G.H.
\item \textbf{Data availability}. The data that support the findings of this study are available on request from the corresponding author. 

\end{itemize}

\newpage

\newpage
\newpage
\newpage
\newpage
\newpage
\clearpage

\begin{appendices}

\section*{Supplementary Figures}

\begin{figure}[h!]
    \centering
    \makebox[0pt]{\includegraphics[width=1.37\linewidth]{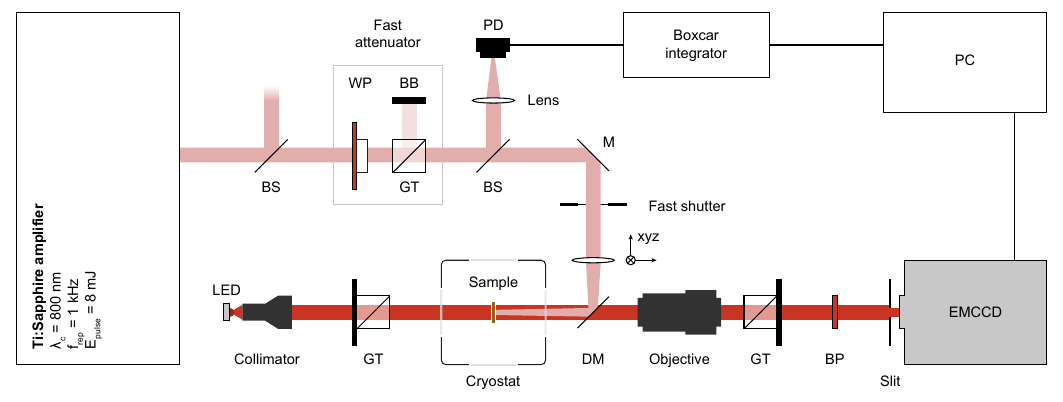}}
    \caption{\textbf{$\mid$ S1: Experimental setup.} To image the dynamics of the wFM order in DTFO during rapid thermal quenches, the output of a LED is collimated and polarized by a Glan-Taylor prism (GT). The probe beam is guided into the optical cryostat, transmitted through the sample, and collected by a camera objective. The image contrast is generated by adjusting a second Glan-Taylor prism (GT) to filter the rotated polarization component. The light passes a band pass filter (BP) and passes an adjustable slit array used to mask selected parts of the detector (EMCCD) for imaging at enhanced frame rates. Sample excitation is realized by focusing the output of a Ti:Sapphire amplifier onto the sample by a plano-convex lens. The beam can be attenuated with high precision and speed using the combination of a half-wave plate (WP) and a Glan-Taylor prism (GT). BB, beam block. A glass slide (BS) is used to redirect a small fraction of the attenuated beam onto a photodiode (PD) to record a reference signal proportional to the sample excitation during heating, annealing and quenching. A fast shutter can be used to create even higher gradients in the time-dependent optical excitation profile. The pump beam is guided onto the back-side of the sample via a dichroic mirror (DM), which is highly reflective in the wavelength range of the pump, and highly transmissive at the probe wavelength of 650\,nm.}
    \label{fig:enter-label}
\end{figure}
\newpage
\newpage
\begin{figure}[ht!]
    \centering
    \makebox[0pt]{\includegraphics[width=1.37\linewidth]{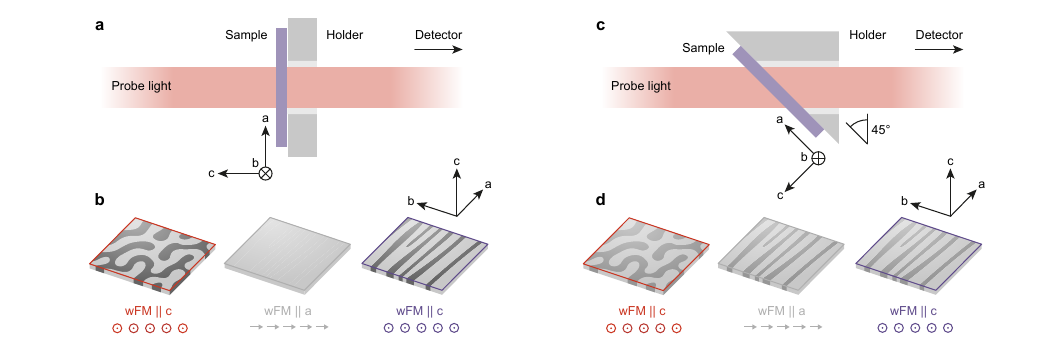}}
    \caption{\textbf{$\mid$ S3: Measurement geometries.} \textbf{a}, Sketch of the sample orientation for the measurements presented in Fig.\,\ref{fig:2} and Fig.\,\ref{fig:3}. \textbf{b}, Schematic depiction of the image contrast in $F_z$ and $F_x$ phases for vertical sample orientation. \textbf{c}, Sketch of the sample orientation for the measurements presented in Fig.\,\ref{fig:1} and Fig.\,\ref{fig:4}. \textbf{d}, Schematic depiction of the image contrast in $F_z$ and $F_x$ phases for tilted samples.}
    \label{fig:S3}
\end{figure}

\newpage
\begin{figure}[ht!]
    \centering
    \makebox[0pt]{\includegraphics[width=1.37\linewidth]{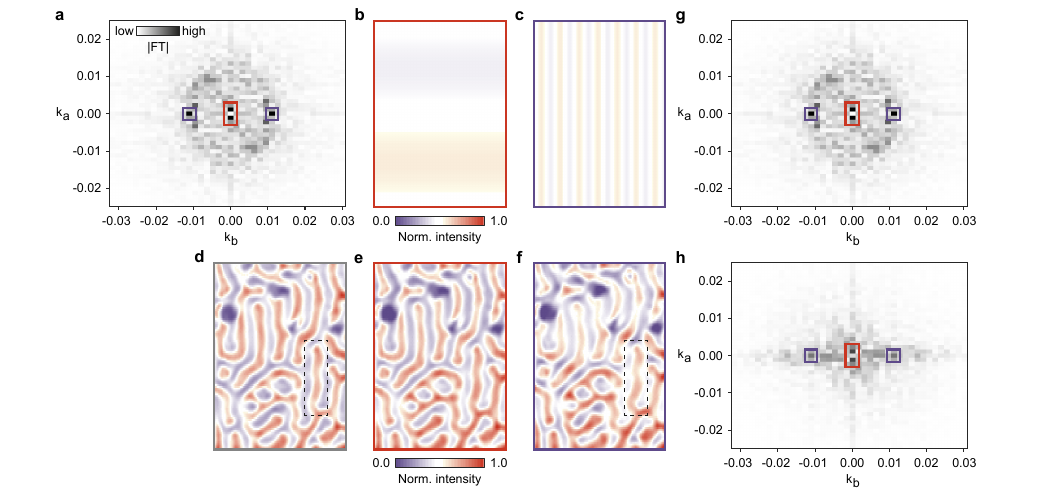}}
    \caption{\textbf{$\mid$ S4: Supplement to the Fourier analysis in Fig.\,4.} \textbf{a}, 2D Fourier transform of the initial $\text{HT-}F_z$ domain pattern taken from the dataset underlying Fig.\,\ref{fig:4}b-d. Red and violet rectangles highlight two types on Fourier components dominating the 2D FT around $k_{\text{a}}=0$ (see also the prominent feature at $k_{\text{a}}=0$ before $t=0\,\text{ms}$ in Fig.\,\ref{fig:4}b). \textbf{b}, Real space pattern yielded by an inverse FT of the two prominent Fourier components (red) in \textbf{a}. These features in the FT correspond to a vertical intensity gradient across the entire image caused by a slightly inhomogeneous illumination of the sample. \textbf{c}, Real space pattern yielded by an inverse FT of the two prominent Fourier components (violet) in \textbf{a}. We associate this feature with the general anisotropy of the system, which is present in all phases, including the $\text{HT-}F_z$ phase. In the case of the $F_x$ and $\text{LT-}F_z$ phases, this anisotropy leads to the formation of stripes along the $a$-direction. For the $\text{HT-}F_z$ phase, other interactions dominate domain formation. However, the characteristic maze domains of the $\text{HT-}F_z$ do exhibit a slight tendency to align along the $a$-direction resulting in peaks at $(k_a=0,k_b=\pm0.011\,\mu\text{m}^{-1})$ in the FT. \textbf{d}, Original domain image. \textbf{e}, Domain image yielded by an inverse FT of \textbf{a} after filtering out the components highlighted in red. Notice the smaller intensity gradient along the vertical direction in \textbf{e} compared to \textbf{d}. \textbf{f}, Domain image yielded by an inverse FT of \textbf{a} after filtering out the components highlighted in violet. Compare the slightly larger modulation of domains along $b$ highlighted by the black dashed rectangles in \textbf{d} and \textbf{f}. Overall, the highlighted features in Fourier space do not dominate the domain pattern and are therefore not relevant for the analysis presented in the main text. \textbf{g}-\textbf{h}, FTs of domain images before (\textbf{g}) and shortly after (\textbf{h}) SRT1. Whereas features associated with the intensity gradient (red) are present throughout the entire duration of the measurement, features corresponding to the systems anisotropy (violet) are transiently suppressed following SRT1.}
    \label{fig:S4}
\end{figure}

\newpage
\end{appendices}



\end{document}